\documentclass[twocolumn,aps,prl,amssymb,amsmath,longbibliography,superscriptaddress]{revtex4-2}

\usepackage{graphicx}% Include figure files
\usepackage{dcolumn}% Align table columns on decimal point
\usepackage{bm}% bold math
\usepackage{nicefrac} 
\usepackage[usenames,dvipsnames]{xcolor}
\usepackage{ulem}
\usepackage[separate-uncertainty=true,multi-part-units=single]{siunitx}

\newcommand{\MW}[1]{\textcolor{olive}{#1}}

\begin{document}

\title{Non-Equilibrium Multiplet Excitations probed by the $M_{5,4}$ Branching Ratio in $3d \rightarrow 4f$ X-ray Absorption Spectroscopy}

\author{Tim Amrhein}
\email{timamrhein@zedat.fu-berlin.de}
\affiliation{Fachbereich Physik, Freie Universit{\"a}t Berlin, Arnimallee 14,
	14195 Berlin, Germany}
    
\author{Beyza Salantur}
\affiliation{Fachbereich Physik, Freie Universit{\"a}t Berlin, Arnimallee 14,
	14195 Berlin, Germany}
    
\author{Ralph P{\"u}ttner}
\affiliation{Fachbereich Physik, Freie Universit{\"a}t Berlin, Arnimallee 14,	14195 Berlin, Germany}    
\author{Niko Pontius}
\affiliation{Helmholtz-Zentrum Berlin für Materialien und Energie GmbH, Albert-Einstein-Str. 15, 12489 Berlin, Germany.}
\author{Karsten Holldack}
\affiliation{Helmholtz-Zentrum Berlin für Materialien und Energie GmbH, Albert-Einstein-Str. 15, 12489 Berlin, Germany.}
\author{Ru-Pan Wang}
\affiliation{Deutsches Elektronen-Synchrotron DESY, Notkestraße 85, 22607 Hamburg, Germany.}
\author{Christian Sch{\"u}{\ss}ler-Langeheine}
\affiliation{Helmholtz-Zentrum Berlin für Materialien und Energie GmbH, Albert-Einstein-Str. 15, 12489 Berlin, Germany.}
\author{Martin Weinelt}
\affiliation{Fachbereich Physik, Freie Universit{\"a}t Berlin, Arnimallee 14,	14195 Berlin, Germany}
\author{Nele Thielemann-K{\"u}hn}
\affiliation{Fachbereich Physik, Freie Universit{\"a}t Berlin, Arnimallee 14,	14195 Berlin, Germany}

\date{\today}% It is always \today, today,

\begin{abstract}

We show that ultrafast electronic $4f$ multiplet transitions in terbium metal are manifested by changes in the relative spectral weight of the $M_5$ and $M_4$ X-ray absorption resonances. Our experimental results are supported by a simulation of excited multiplet spectra with atomistic calculations; they prove that the so-called third rule of Thole and van der Laan, which relates the branching ratio of the spin-orbit split resonances to the total angular momentum $J$ of the excited ion, is also valid in non-equilibrium. The presented detection scheme allows to detect $J$-changing excitation, \textit{i.e}, alterations of spin and orbital states, even in samples without net magnetization. This makes branching-ratio spectroscopy a powerful tool for the quantitative investigation of ultrafast changes in angular momentum $J$.

\end{abstract}

\maketitle{} 

In the field of ultrafast spin dynamics research, one main goal is to identify and exploit mechanisms that transfer energy and angular momentum between microscopic subsystems, thereby allowing us to efficiently control magnetization dynamics on the femtosecond time scale.
The coupling between orbital, spin, and lattice subsystems, which govern the magnetic order, is uniquely determined by the electronic states. 
Electronic excitations can in turn alter intrinsic coupling parameters with the potential to form new material phases on ultrashort timescales.
They further promise energy-efficient functionalities operating with minimal energy consumption, as thermal occupation of the electronic system typically requires an order of magnitude less energy than equivalent lattice heating \cite{Anisimov1974,Allen1987,Koopmanns2010}.
In both, magnetically ordered transition metal (TM) compounds and rare-earth (RE) systems, electronic excitations have been shown to directly impact the spin dynamics \cite{Willems2020,NTK2024,Pontius2022}.
For example, laser-induced electronic excitations in $3d$ TM alloys are discussed to induce optical inter-site spin transfer (OISTR) between different atomic sites \cite{Willems2020}, as well as to enhance spin-flip scattering of majority electrons into laser generated minority holes through increased spin-orbit coupling \cite{Pontius2022}, both having immediate impact on the magnetization  dynamics. 
In RE orthoferrites, optically driven $4f$ multiplet excitations affect the magneto-crystalline anisotropy, reorienting the Fe $3d$ spins \cite{Kimel2004,Mikhaylovskiy2015,Baierl2016,Schlauderer2019,Fitzky2021}. 
In the case of RE $4f$ metals direct optical $f - f$ excitations are suppressed, but it has been recently shown, that inelastic $5d - 4f$ electron-electron scattering provokes multiplet transitions in the $4f$ shell \cite{NTK2024}.
These excitations change the total angular momentum $J$, \textit{i.e.}, the orbital and spin wave-functions, and thereby alter $4f$-spin-lattice coupling, known to govern intrinsic magnetization dynamics in RE metals \cite{Frietsch2020}.\\

Although the role of electronic excitations receives growing attention in researches on ultrafast spin dynamics \cite{Dewhurst2018,Harris-Lee2024}, their experimental proof remains challenging \cite{Siegrist2019,Hofherr2020,Willems2020,Golias2021,Bobowski2024}. 
Time-resolved resonant inelastic X-ray scattering (RIXS)  provides a unique state- and element-selective probe of electronic structure dynamics \cite{NTK2024,Mitrano2024,Groot2024, Baker2017, Glatzel2005, Kotani2001}. 
However, its feasibility demands a high X-ray photon flux as RIXS cross-sections in the soft X-ray regime are intrinsically small.
With the required femtosecond time resolution, these experiments become currently accessible at Free Electron Laser (FEL) sources only.
Electronic structure dynamics can also be probed via changes of the absorption multiplet shape in X-ray absorption spectroscopy (XAS) \cite{NTK2024}, but these experiments require high energy resolution to resolve the rather subtle changes.
Here we show a complementary approach for probing electronic excitations by X-ray absorption spectroscopy that does not require high energy resolution and is, therefore, well suited for high-bandwidth sources of X-ray pulses: 
Electronic $4f$ excitations in Tb manifest as a change of the branching ratio of the $M_5$ and $M_4$ XAS resonances, which according to the third rule of Thole and van der Laan is determined by their specific total angular momentum  $J$ \cite{GvanderLaan1986}. 
This experimental approach allows us to track ultrafast multiplet excitations, even at low photon fluxes and with limited photon energy resolution by evaluating integrated XAS intensity changes of spin-orbit split resonances.
As the presented method is not based on magnetic contrast, changes in the spin and orbital states that modify the atomic angular momentum $J$ can be studies even in non-magnetic samples.\\

\begin{figure*}
	\includegraphics[width=1.8\columnwidth]{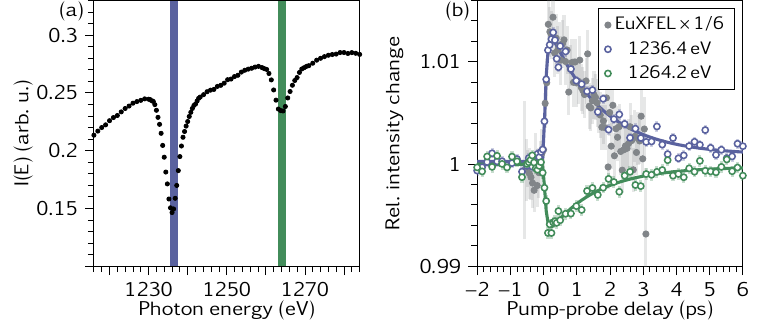}
	\caption[transmission data]{Experimental Tb $M_{5,4}$ transmission data. (a) Static $M_{5,4}$ transmission spectrum. Vertical blue and green bars mark the energy range for which the delay traces in (b) where recorded.  (b) Transient relative change of transmitted intensity at $M_5$ (\SI{1236.2}{\eV}, blue circles) and $M_4$ (\SI{1264.2}{\eV}, green circles) edges recorded at FemtoSlicing. Solid blue and green lines are bi-exponential least square fits to the data (see SM). For comparison, the gray markers show the relative change of the transmitted signal recorded at the Tb $M_5$ resonance in the high-energy resolution XAS experiment performed at EuXFEL. The latter data were scaled by a factor of $\nicefrac{1}{6}$ to account for the different resolution and pump fluence. Error bars correspond to the error-propagated standard deviation $\sigma$. \\
    }
 
\label{fig:trans}
\end{figure*}

In $4f$ RE metals XAS allows probing dipole transitions from $3d$ core- to localized $4f$ valence-levels. 
In $3d$ transition metals the corresponding excitation into the itinerant $3d$ valence-orbitals occurs from $2p$ core levels. 
Well spin-orbit split $3d_{\nicefrac{5}{2}}$ and $3d_{\nicefrac{3}{2}}$ core levels give rise to  $M_{5}$ and $M_{4}$ absorption edges of the RE, while $2p_{\nicefrac{3}{2}}$ and $2p_{\nicefrac{1}{2}}$ components correspond to the $L_{3}$ and $L_{2}$ absorption edges in $3d$ metals, respectively \cite{stohrMagnetism2006}.   
The strength of these resonances is directly related to the number of valence holes as described by the charge sum-rule \cite{stohrMagnetism2006}: The integrated intensity of both spin-orbit split components $(I_{c+s}+I_{c-s})$ is proportional to the number of holes in the valence states of the ground-state. 
Here, $s$ denotes the spin quantum number and $c$ is the orbital angular momentum of the core shell.
The branching ratio $B = I_{c+s}/(I_{c+s}+I_{c-s})$ quantifies how the intensity splits between both components.  
$B$ does  only in first approximation correspond to the ratio given by the number of electrons in the spin-orbit split core levels $B^0$, where $B^0 = 6/(6 + 4) = 3/5$ for RE and $B^0 = 4/(4 + 2) = 2/3$ for TM.
The deviation from this simple ratio is directly connected to specific electrostatic interactions between core- and valence electrons in the XAS final state.
Atomistic calculations predict that $B$ increases gradually with the total angular momentum $J$ of the valence electrons.
Accordingly, for a more than half filled shell, the highest branching ratio is observed for the Hund's ground-state with the largest $J$ (see Supplemental Material, SM).
This relation is expressed in the so-called 3rd rule of Thole and van der Laan \cite{PThole1988}. 
It has been experimentally confirmed in the static case for different electronic configurations along the series of pure $3d$ TMs \cite{Waddington1986} as well as for the same elements embedded in various compounds \cite{GvanderLaan1986} with different degree of hybridization and resulting variation of $J$.\\

We show that the 3rd rule of Thole and van der Laan remains valid even for excited states in the non-equilibrium, which provides a means to characterize changes of the total angular momentum $J$ by time-resolved XAS.
To this end we chose metallic Tb.
Recent time-resolved FEL experiments with high photon-energy resolution (XAS and RIXS) on Tb metal revealed $4f$ excitations out of the $^{7}F_6$ ground-state primarily into the $^7F_5$ excited state. 
These multiplet transitions are driven by inelastic electron-electron scattering between optically excited (hot) $5d$ valence electrons and (cold) localized $4f$ states \cite{NTK2024}.
In the present study, we identify the $^{7}F_6 \rightarrow\,^{7}F_5$ transitions involving $\Delta J=-1$ via direct signatures in the X-ray absorption $M_{5,4}$ branching ratio $B$.
The experimental time resolution of 130\,fs allows us to resolve the temporal evolution of the Tb $4f$ electronic excitations. 
The latter follow the temperature of the laser-heated $5d6s$ valence electronic system and hence appear on a 100\,fs time scale and decay again within a few ps.\\

 \begin{figure*}
	\includegraphics[width=1.8\columnwidth]{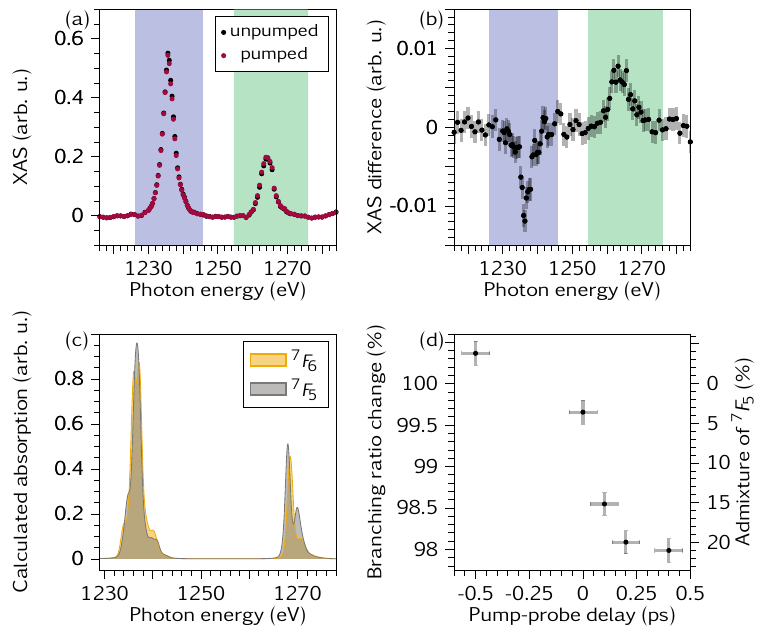}
	\caption[branching ratio]{Quantitative evaluation of branching ratio (a) Background-corrected Tb $M_{5,4}$ XAS spectra for the unpumped sample (black dots) and for the sample excited with an 800-nm-laser pulse, \SI{400}{\femto\s} after pump-pulse arrival (red dots). (b) Absolute difference between the two spectra shown in (a). The colored areas in (a) and (b) indicate the $M_5$ and $M_4$ integration region, uses for the calculation of the branching ratio. Error bars in (a) and (b) correspond to the error-propagated standard deviation. (c) Atomistically calculated spectra for the ground-state $^7F_6$ (yellow shaded area) and for the first excited state $^7F_5$ (gray shaded area). (d) Pump-induced relative change of the branching ratio with respect to the value for the unpumped sample (left ordinate). The branching ratio decreases by about \SI{2.0\pm0.2}{\percent} within the first \SI{400}{\femto\s}. Correspondingly \SI{21\pm2}{\percent} of the ions in the probed volume are excited (right ordinate). Error bars correspond to the error-propagated standard deviation $\sigma$.\\}
	\label{fig:BR}
\end{figure*}

The time-resolved XAS measurements have been conducted in transmission geometry using the DynaMaX instrument at the Femtoslicing facility of BESSY\,II \cite{Holldack2014}. This storage-ring-based ultrafast soft X-ray source provides a time resolution of 130\,fs along with a moderate energy resolving-power of E/$\Delta$E$\leq$500.
The sample was a Tb thin film (\SI{12}{\nano\m}) grown on an Al~foil (\SI{100}{\nano\m}) with Y buffer and capping layers (\SI{40} and \SI{3}{\nano\m}) for oxidation protection (see SM). 
We investigated the $M_{5,4}$ resonances at room temperature in the paramagnetic phase of Tb and analyzed their branching ratio before and after excitation. 
For sample excitation, 800-nm laser pulses of \SI{6\pm2}{\milli\J/\centi\m\squared} incident laser fluence were used.
During data accumulation we corrected for possible monochromator photon energy drifts and variations of the delay between optical pump and X-ray probe pulses by repeated reference measurements (see SM).\\

Figure\,\ref{fig:trans}a shows the static transmitted intensity, recorded over the Tb $M_{5,4}$ resonance. 
The vertical bars in Fig.\,\ref{fig:trans}a mark the photon energy that has been used to track pump-induced intensity changes at the $M_5$ (\SI{1236.4}{\eV}) and $M_4$ (\SI{1236.4}{\eV}) edges, respectively.
As depicted in Fig.\,\ref{fig:trans}b, the $M_5$ and $M_4$ edge intensities show opposite dynamics: The $M_5$ absorption is transiently reduced (increased transmission), while the $M_4$ absorption increases (reduced transmission).
The pump-effect at $M_5$ and $M_4$ edges directly involves a change of the branching ratio $B$.
With a response time of about \SI{80\pm10}{\femto\s} and a recovery time constant of \SI{1.5\pm0.2}{\pico\s} (see fits in Fig.\,\ref{fig:trans}b and SM), the transient changes follow the same temporal characteristics as observed in the high-energy resolution XAS and RIXS experiment \cite{NTK2024}. 
For direct comparison, Fig.\,\ref{fig:trans}b includes the relative change of transmitted intensity, as recorded at the $M_5$ resonance in the XAS-experiment at EuXFEL.
The latter has been directly related to the 4$f$ multiplet excitations, following the temperature rise and fall of the optically excited $5d$ valence electrons.
Please note that the 800-nm pump pulse (1.55-eV photon energy) cannot excite the $4f$ system directly.
We consequently assign the $M_{5,4}$ intensity and related branching ratio changes to variations of the total $4f$ angular momentum $J$, originating from $5d-4f$ inelastic electron-electron scattering.\\

For a more quantitative evaluation of the 4$f$ multiplet excitations, we recorded the full $M_{5,4}$ resonance for different pump-probe delays and modeled changes of the branching ratio with the Tb $M_{5,4}$ X-ray absorption spectra from atomistic calculations presented in Ref.\,\cite{NTK2024}. 
The raw X-ray signal measured in the experiment is the sample transmitted X-ray intensity depicted in Fig.\ref{fig:trans}a. 
This signal does not only comprise the absorption by the Tb $M_5$ and $M_4$ resonances, but also absorption from transitions into other (mainly $p$-like) final valence states giving rise to edge-jumps, \textit{i.e.}, step-like increases of the absorption within the $M_5$ and $M_4$ resonance. 
These, together with non-resonant spectral contributions of the sample stack, cause the background in the XAS signal.
Moreover the varying beamline transmission and detector sensitivity result in long-range intensity modulations across the transmission spectrum. 
With the X-ray absorption determined as the natural logarithm of the transmitted intensity, we can correct for these `background` contributions by rescaling the raw X-ray transmission signal with a model background approximated from the spectral regions outside the resonances (see SM). 
The resulting background-corrected XAS signal is directly proportional to the absorption coefficient.
Such processed XAS spectra are shown in Fig.\,\ref{fig:BR}a for the unpumped and excited sample, recorded at a pump-probe delay of 400\,fs.
Figure\,\ref{fig:BR}b illustrates the absolute difference between the two XAS spectra in Fig.\,\ref{fig:BR}a and reveals a robust pump-effect, well detectable within the noise level.
The branching ratio of the ground-state spectrum in Fig.\,\ref{fig:BR}a is directly comparable to the one given by the atomistic calculation. 
Based on the integrals $I_{M_{5}}$ and $I_{M_{4}}$ over the energy regions marked in Fig\MW{s}.\,\ref{fig:BR}a and b, the absolute experimental branching ratio \mbox{$B_{\rm{exp}}=I_{M_{5}}/(I_{M_{5}}+I_{M_{4}})$} is \SI{0.73\pm0.02}{} for the unpumped sample.
This fairly well matches the value of about $B_{^7F_6}=0.76$ given by the atomistic $^7F_6$ ground-state spectrum, as shown in Fig.\,\ref{fig:BR}c.
The experimental error is deduced from variations of the branching ratio, which was determined from unpumped spectra recorded as reference at each pump-probe delay.
For a quantitative analysis of the 4$f$ excitations, we evaluate the \textit{relative} branching-ratio change for given delay, \textit{i.e.}, the ratio between $B_{\rm{exp}}$ for pumped and unpumped signal. 
The latter can be deduced with high accuracy, since pumped and unpumped signals are acquired in a shot-to-shot alternating manner at a 3\,kHz repetition rate providing excellent normalization \cite{Holldack2014,Schick2016}.
In this way, we find that the experimental branching ratio decreases by about \SI{2.0\pm0.2}{\percent} within the first \SI{400}{\femto\s} (Fig.\,\ref{fig:BR}d).
The branching ratio determined for different superpositions of the atomistic $^7F_6$ and $^7F_5$ state spectra (Fig.\,\ref{fig:BR}c) yields a maximum change of \SI{10}{\percent} for the pure $^7F_5$ state with respect to the ground-state.
Accordingly, the \SI{2.0\pm0.2}{\percent} relative change of the experimental branching ratio, correspond to \SI{21\pm2}{\percent} excited $^7F_5$ ions in the probed volume (Fig.\,\ref{fig:BR}d, right ordinate). 
The residual \SI{79\pm2}{\percent} remain in the ground-state $^7F_6$.\\

Our analysis considers excitations into the $^7F_5$ state, which were unambiguously identified by the previous high-energy resolution XAS and RIXS experiment \cite{NTK2024}. 
We have neglected possible but minor contributions from energetically higher lying states of the multiplets. 
Our results are in line with earlier observations made for a comparable absorbed energy density in the Tb layer (see SM), which suggest about \qtyrange[range-phrase=~--~,range-units = single]{16}{20}{\percent} of the probed ions to be $^7F_5$ excited.\\

We note that atomistic calculations and predicted branching ratios are generally dependent on Slater-Condon and crystal-field parameters. 
The former describe screening of $3d-4f$ and $4f-4f$ interactions, the latter the influence of the neighboring ions' electrostatic field on the $4f$ wave-function. 
Since in $4f$ metals spin-orbit coupling dominates over the crystal-field interaction, calculations are based on spherical wave functions, neglecting the crystal field. 
However, Slater-Condon parameters may change by alterations of electronic screening. 
The \SI{1.55}{\eV} pump pulse only excites electrons within the $5d6s$ valence band; $5d\rightarrow 4f$ and $4f \rightarrow 5d$ optical transitions would require higher energies of \qtyrange[range-phrase=~and~,range-units = single]{2.8}{2.3}{\eV}, respectively \cite{Lang1981}.
One could still speculate that electronic screening changes caused by $5d$-excitations affect the branching-ratio.  
However, the time-resolved RIXS data in Ref. \onlinecite{NTK2024} do not show any transient energy shifts of the multiplet lines after laser excitation. 
This excludes pump-induced alterations of the electronic screening to affect $3d-4f$ and/or $4f-4f$ interactions in the $4f$ metals. 
It is the metallic ground-state and the localized character of the $4f$ states, which allows us to reliably simulate the absorption multiplets, even for optically excited samples \cite{Groot2008}. 
We can, therefore, safely assign the experimental branching ratio variations to be distinctly driven by 4$f$ multiplet excitations changing $J$.\\

In summary, we examined ultrafast changes in the branching ratio of the Tb $M_{5,4}$ absorption resonance after near-infrared laser excitation. Branching-ratio spectroscopy is shown to be highly sensitive to $J$ changing $4f$ multiplet excitations.
We could quantify the amount of $^7F_6\,\rightarrow\,^7F_5$ excited ions in the probed volume, demonstrating the 3rd rule from Thole and van der Laan to be applicable to excited states on ultrashort timescales. The presented experimental approach can be exploited at instruments of limited energy resolution or as an additional tool in high-energy-resolved XAS.
High-throughput studies, aiming for quick quantification of branching-ratio active multiplet excitation, might become possible, \textit{e.g.}, in setups using a broad undulator harmonics without monochromator.
New insights into ultrafast magnetization dynamics, in particular, will be achieved by time-resolved detection of branching ratio and XMCD, relating electronic transitions with concurrently altered magnetic order.\\

\begin{acknowledgments}
This work was supported by the Deutsche Forschungsgemeinschaft through TRR~227 "Ultrafast Spin Dynamics", ProjectID 328545488 (projects A01 and A03) and the Bundesministerium für Bildung und Forschung, Project "Spinflash" (05K22KE2). 
The measurements were carried out at the Femtoslicing Facility of the BESSY II electron storage ring operated by the Helmholtz-Zentrum Berlin für Materialien und Energie. The Tb samples were prepared in the DynaMaX sample-preparation chamber and statically characterized at the PM3 scattering instrument. 

\end{acknowledgments}

%\nocite{*}

\bibliography{manuscript}
% Produces the bibliography via BibTeX.

\end{document}

\preprint{APS/123-QED}

\title{Supplemental Material \\
Non-Equilibrium Multiplet Excitations probed by the $M_{5,4}$ Branching Ratio in $3d \rightarrow 4f$ X-ray Absorption Spectroscopy}

\author{Tim Amrhein$^1$}
\email{timamrhein@zedat.fu-berlin.de}
\author{Beyza Salantur$^1$}
\author{Ralph P{\"u}ttner$^1$}
\author{Niko Pontius$^2$}
\author{Karsten Holldack$^2$}
\author{Christian Sch{\"u}{\ss}ler-Langeheine$^2$}
\author{Martin Weinelt$^1$}
\author{Nele Thielemann-K{\"u}hn$^1$}

\affiliation{
$^1$ Freie Universität Berlin, Fachbereich Physik, Arnimallee 14, 14195 Berlin, Germany.
}
\affiliation{
$^2$ Helmholtz-Zentrum Berlin für Materialien und Energie GmbH, Albert-Einstein-Str. 15, 12489 Berlin, Germany.
}

\date{\today}% It is always \today, today,

\maketitle{}

\subsection{Sample preparation}
The Tb thin-film transmission sample was prepared by molecular beam epitaxy employing home-build evaporators in the DynaMaX sample-preparation chamber at the PM3 beamline of BESSY II (Helmholtz-Zentrum Berlin). 
The substrate was a \SI{100}{\nano\m} Al foil on a Cu frame. 
A buffer layer of \SI{40}{\nano\m} Y was grown on top of the substrate with a growth rate of \SI{2}{\angstrom\per\min}. The buffer layer protects the Tb film against contamination originating from the substrate and reduces the strain in the Tb film. 
On top of the buffer layer, a \SI{12}{\nano\m} Tb film was grown with a growth rate of \SI{6}{\angstrom\per\min}. 
To protect the Tb film from oxidation, the sample was capped with another \SI{3}{\nano\m} of Y grown at a rate of \SI{3}{\angstrom\per\min}. The base pressure of the chamber was \SI{5e-10}{\milli\bar} and increased to \SI{1e-9}{\milli\bar} during the growth process.\\

\subsection{Background correction of Tb $M_{5,4}$ transmission spectra}
The background of the transmission spectra was modeled with a polynomial function of second order, combined with two Gau\ss{}ian-broadened heavy-side functions, taking into account the edge jumps at the $M_5$ and $M_4$ resonance, respectively.
We used  the following model within the lmfit routine in Python: \\

\begin{lstlisting}
def backgroundEJ(x,s1,s2,offX,offY,be1,lt1,sc1,be2,lt2,sc2): 
    model=(offY-s2*(x-offX)**2
        -s1*(x-offX)-edge_jump(x,be1,lt1,sc1)
        -edge_jump(x,be2,lt2,sc2))
    return model
\end{lstlisting}
with
\begin{lstlisting}
from PyAstronomy import pyasl
def edge_jump(x,be,lt,sc):
    E=linspace(-30,60,num=7001,endpoint=True)
    edgejump=pyasl.broadGaussFast(E, sc*heaviside(E-be,0), lt, 
        edgeHandling=None, maxsig=None)# broadening
    Theo_EJ=[]
    for j in range(0,len(x)):
        Theo_EJ.append(interp(x[j],E,edgejump)) 
    return Theo_EJ
\end{lstlisting}

\begin{figure}
	\centering
	\includegraphics[]{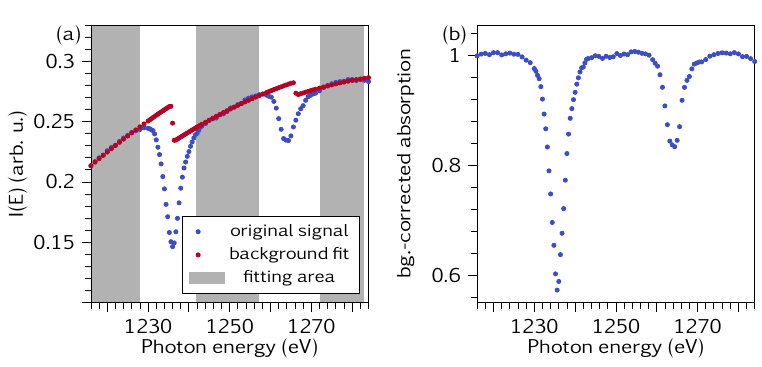}
	\caption[background correction]{a) X-ray transmission spectrum recorded at the Tb $M_{5,4}$ resonance\MW{s}. While the blue dots display the experimental data, the red dots show the polynomial fit of second order to the background, including the edge jumps. Only the gray marked regions were used for fitting. b) Background corrected transmission spectrum obtained by normalizing of the experimental signal to the fitted background.}
	\label{fig:7F5theory}
\end{figure}

The amplitude of the linear and quadratic part, is given by s1 and s2, respectively.
offY is a constant contribution to the polynom, while offX introduces an offset in energy\,x.
be1 and be2 are the energy positions of the two edge jumps and sc1 and sc2 correspond to the respective amplitudes. 
lt1 and lt2 define the energy broadening of the edge jumps and are set to lt1=lt2=0.2\,eV, according to the estimated core-hole lifetime broadening. 
The edge-jump energy positions are set with respect to the energy of the $M_5$ resonance maximum and are be1=0\,eV for the $M_5$ edge and be2=30\,eV for the $M_4$ edge .
The resulting parameters from the background fits to the spectra, recorded at different pump-probe delays, are listed in Tab. \ref{tab:fit_para_bg}.
Fig. \ref{fig:7F5theory}a shows the transmission spectrum for the unpumped sample, obtained together with the spectrum at 0.4\,ps pump-probe delay (blue dots). 
For the background fit (red dots), only the energy regions outside the absorption edges were considered (marked in gray in Fig. \ref{fig:7F5theory}a).
The background corrected transmission spectrum (Fig. \ref{fig:7F5theory}b) is obtained by division of the spectrum in Fig. \ref{fig:7F5theory}a by the background. 
The corresponding XAS signal, shown in Fig.\,2a of the main text, corresponds to the negative natural logarithm of the background corrected transmitted intensity.\\
 
\begin{table}[h]
    \centering
    \caption[$M_{5,4}$ XAS background fit parameters]{Fit parameters used for fitting the background of the Tb transmission around the $M_5$ and $M_4$ resonance for different pump-probe delays. For the alternatingly recorded spectra for pumped and unpumped sample, the same background is subtracted. \vspace{6px}}			
    \label{tab:fit_para_bg}
    \begin{tabular}{|c|c|c|c|c|c|c|}
        \hline
         &s1&s2&offX&offY&sc1&sc2\\
         \SI{-0.5}{\pico\s}&$4\cdot10^{-9}$&$1.3\cdot10^{-6}$&81&0.036&0.001&0.003\\
         \SI{0}{\pico\s}&$5\cdot10^{-5}$&$1.7\cdot10^{-6}$&78&0.033&0.001&0.003\\
         \SI{0.1}{\pico\s}&$1.3\cdot10^{-5}$&$2.0\cdot10^{-6}$&53&0.029&0.001&0.003\\
         \SI{0.2}{\pico\s}&$7\cdot10^{-6}$&$2.3\cdot10^{-6}$&50&0.033&0.001&0.002\\
         \SI{0.4}{\pico\s}&$5\cdot10^{-10}$&$1.4\cdot10^{-6}$&70&0.032&0.002&0.002\\
        \hline	
    \end{tabular}
\end{table}

\subsection{Absorbed energy density}
The incident angle of the nearly coaxial optical pump and X-ray probe beam was 42\,degrees with respect to the sample normal.
Considering the measured beam power, spot size, and 3 kHz repetition rate of the pump pulses, we estimate an incident pump fluence of \SI{6\pm2}{\milli\J/\centi\m\squared}.
We calculated the absorbed energy density to be \SI{1.0\pm0.4}{\kilo\J/\centi\m\cubed} in the Tb layer, by using the program IMD by David Windt \cite{Windt1998}.
Within the uncertainties, this is similar to the absorbed energy density of \SI{2.4\pm1.3}{\kilo\J/\centi\m\cubed}, which was determined for the experiment at XFEL \cite{NTK2024}.

\subsection{Fit function for the temporal profile of pump-probe delay traces}
The temporal evolution of the pump effect shown in Fig.\,1b of the main text was fitted with the following model using the lmfit routine in Python.\\
\begin{lstlisting}
def doubleDecayConvRec(x, t0, tau1, tau2, A, C, sigma, offs):
	term1 = expConvGauss(x-t0,tau1,A,sigma) 
	term2 = expConvGauss(x-t0,tau2,-C,sigma)
	term3 = ABCHConvGauss(x-t0,A,C,sigma)
	return (term1 + term2 + term3 + offs)
\end{lstlisting}
with
\begin{lstlisting}
def expConvGauss(x,tau,A,sigma): 
	term1 = np.exp(-x/tau)*np.exp(sigma**2/(2*tau**2))*
		(special.erf((sigma**2-x*tau)/
                (np.sqrt(2)*sigma*tau))-1)
	return -A/2*(term1)
\end{lstlisting}
and
\begin{lstlisting}
def ABCHConvGauss(x,A,C,sigma):
	term1 = -A/2*special.erf(x/(np.sqrt(2)*sigma))
	term2 = +C/2*special.erf(x/(np.sqrt(2)*sigma))
	return term1+term2+1-A/2+C/2
\end{lstlisting}
In this model t0 fits the start of the exponential decay. 
tau1 is the time constant of the exponential rise/drop with amplitude A. 
tau2 is the recovery time constant with an amplitude C. 
The temporal broadening sigma was calculated from the temporal resolution $0.13/(2*\sqrt{2*\ln(2)})$ and kept constant. The parameter offs is used to align the base level of the fit and the data.

\subsection{Data accumulation}
Despite the FemtoSlicing facility providing stable experimental conditions regarding photon energy and pump-probe overlap, drifts might occur during measurement over several days.\\
To correct for drifts in the photon energy, the pronounced Tb $M_5$ resonance was measured every two hours and fitted via a Gau\ss{}ian profile to evaluate the absorption maximum of the sample.\\
The timing drifts between pump and probe pulses were corrected for by measuring the ultrafast change in absorption at the $M_5$ edge every two hours, fitting it with a double exponential function to determine $t_0$ and shifting the delay accordingly.

\subsection{Branching ratio according to the third rule of Thole and van der Laan}

The third rule of Thole and van der Laan is described in Ref.\ \cite{PThole1988}, however, the 
derivation is presented in detail in Ref.\ \cite{PThole1988b}. 
According to Ref.\ \cite{PThole1988b}, for a free atom in the ground state and under LS coupling conditions the branching ratio can be written as
\begin{equation}
B(\alpha L S J) = B^0 + B^1 = B^0 + A(c,l,n)\cdot Z(\alpha LSJ)
\end{equation}
with 
\begin{equation} \label{eqn_Z}
Z(\alpha LSJ) = \frac{E_{\alpha LSJ} - E_{\alpha LS}}{\xi_l}=
\frac{1}{2} \left( J(J+1) - L(L+1) - S(S+1) \right) \frac{\xi(\alpha LS)}{\xi_l}.
\end{equation}
 Here $L$ is the orbital angular momentum, $S$ the spin momentum and $J$ the total angular momentum while $\alpha$ represents all other quantum numbers. $B^0$ is the statistical branching ratio for the 
spin-orbit component of the core hole with $j = c + \frac{1}{2}$, i.e. 
\begin{equation}
B^0 = \frac{2j+1}{2(2c+1)}.
\end{equation}
$c$ and $l$ are the orbital angular
momentum of the core-hole and the final-state orbital, respectively, and $n$ the number of electrons in the final-state
orbital prior to excitation. $A(c,l,n)$ is a constant tabulated in Ref.\ \cite{PThole1988b}.
Moreover, $E_{\alpha LS}$ is the energy value of the unsplit term and $E_{\alpha LSJ}$ the energy value including 
spin-orbit interaction. Finally, $\xi_l$ is a radial integral commonly referred to as spin-orbit parameter 
(see Ref.\ \cite{Cowan1981} eq.\ 10.9) and $\xi(\alpha LS)$ is the spin-orbit splitting factor.

For $LS$ terms with maximum spin $\frac{\xi(\alpha LS)}{\xi_l}$ can be simplified to
\begin{eqnarray} \label{ls_max}
\frac{\xi(\alpha LS)}{\xi_l} &=& \frac{1}{n} \  {\rm for} \ n < 2l+1 \nonumber \\
&=& 0 \ {\rm for} \  n = 2l+1 \nonumber \\
 &=& \frac{-1}{4l+2-n} \  {\rm for} \ n > 2l+1.
\end{eqnarray}

In the following we shall specify the formulas for the $3d^{10} 4f^8 \rightarrow 3d^9 4f^9$ 
excitations of Tb$^{3+}$. Obviously $c=2$, $l= 3$ and $n=8$ so that $A(d,f,8)=-\frac{2}{45}$ and 
$\frac{\xi(\alpha LS)}{\xi_l} = -\frac{1}{6}$. Moreover $J(J+1) - L(L+1) - S(S+1) = 18$ for $^7F_6$ and 6
for $^7F_5$ so that $B^1 (^7F_6) = \frac{1}{15}$ and $B^1 (^7F_5) = \frac{2}{90}$. From this follows
$B (^7F_6) = \frac{2}{3} \cong 0.667$ and $B (^7F_5) = \frac{56}{90} \cong 0.622$ and finally $B (^7F_5)/B (^7F_6)
= 0.933$. The latter value agrees reasonably well with the calculations presented in Fig. 2 of the manuscript, which
is $B (^7F_5)/B (^7F_6) = 0.904$. Note that the value of $B (^7F_5)/B (^7F_6)= 0.933$ is based on the assumption that
interactions between the core hole and the 4$f$ electrons can be neglected.

\begin{figure}
	\centering
	\includegraphics[]{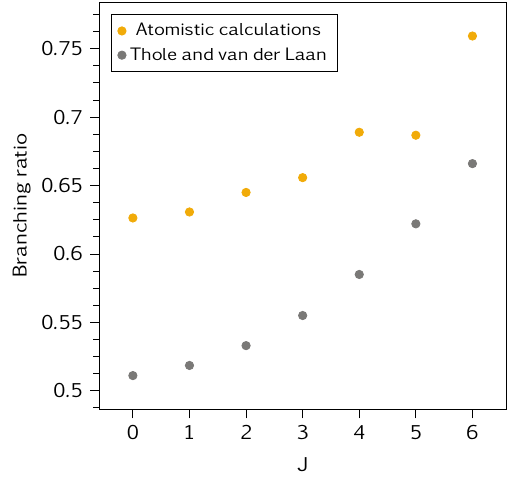}
	\caption[theoretical branching ratio]{Branching ratios of Tb calculated based on the approach of
    Thole and van der Laan (gray dots) as well as based on atomistic calculations (orange dots).}
	\label{fig:BRtheory}
\end{figure}

The obtained branching ratios based on the approach of Thole and van der Laan are displayed in Fig.\ \ref{fig:BRtheory} (gray dots) and are compared with ratios obtained from atomistic calculations for Tb (orange
dots). Both curves show that the branching ratio generally increases with $J$. In more detail, the results based on the approach of Thole and van der Laan show a smooth curve with continuous increase of the difference between the branching rations belonging to two neighboring $J$-values. In contrast to this, the curve based on atomistic calculations are significantly more structured. These differences we refer to the fact that the prerequisites for the approach of Thole and van der Laan are not strictly fulfilled. 

\begin{figure}
	\centering
	\includegraphics[]{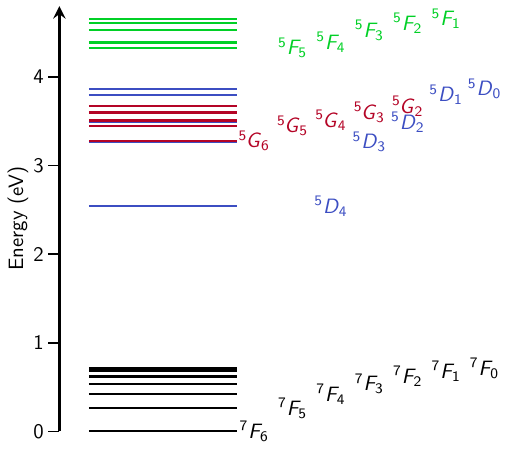}
	\caption[$^7F_J$ multiplet distribution]{Energy levels of the multiplets $^7F_J$, $^5D_J$, $^5G_J$, and $^5F_J$ taken from Ref.\ \cite{Carnall1968}. States with identical $J$ can mix. }
	\label{fig:7FjDistribution}
\end{figure}

In detail, the approach of Thole and van der Laan assumes Russel-Saunders coupling, i.e., the states
are  described well with the term symbols $^{2S+1}L_J$. As a consequence, the states of a given $^{2S+1}L_J$
multiplet should obey the Land\'{e} interval rule, i.e., the energy levels of the different states in 
the multiplet should obey $E_{L,S,J} - E_{L,S,J-1} = aJ $ with $a$ being a constant. It is actually well-known
that for Tb$^{3+}$ the splitting of the $^7F_J$ multiplet  does not strictly fulfill this rule \cite{Carnall1968,Judd1956}. This is due to a violation of the Russel-Saunders coupling, i.e., a mixing of
the $^7F_J$ states with $J$-states of other multiplets. In terms of perturbation theory and by taking into account corrections up to the second-order, this mixing can occur with multiplets where $L$ and $S$ differs by not more than one from the values
of the $^7F_J$ multiplet so that states of the multiplets $^5D_J$, $^5G_J$, and $^5F_J$ can contribute. 
This qualitatively explains the larger branching ratio for the $^7F_4$ state as compared to the neighboring
states $^7F_3$ and $^7F_5$. Due to the relative energetic vicinity of the $^5D_4$, see Fig.\ \ref{fig:7FjDistribution}, a relatively high admixing of this state to the state $^7F_4$ can be expected.
To estimate the influence of the admixing, the $B (^5D_4)$ branching ratio is required. However, since in 
case of the $^5D_J$ multiplet the spin is not maximal, $\frac{\xi(\alpha LS)}{\xi_l}$ cannot be obtained
using eqn.\ \ref{ls_max}. Because of this we estimated $\frac{\xi(\alpha LS)}{\xi_l}$ for the
$^5D_J$ multiplet based on eqn.\ \ref{eqn_Z} and the energy levels calculated by Carnall at al. \cite{Carnall1968} which are shown in Fig.\ \ref{fig:7FjDistribution}. In this way we obtained 
$\frac{\xi(\alpha LS)}{\xi_l}$ for the $^5D_J$ multiplet 
to be by approximately a factor 4 larger than for the $^7F_J$ multiplet. This can also be seen directly in Fig.\ \ref{fig:7FjDistribution} since the energy splitting between two states with consecutive J is significantly larger for the $^5D_J$ multiplet than for the $^7F_J$ multiplet. From the obtained value for 
$\frac{\xi(\alpha LS)}{\xi_l}$ we derived 
 $B (^5D_4)  \cong 0.72$  which is significantly larger than the ratio
$B (^7F_4) = \frac{79}{135} \cong 0.585$ so that an admixing of about 10\% of the state $^5D_4$ to the
state $^7F_4$ can explain an increase of the branching ratio by 0.0135.

\bibliography{manuscript}